\documentclass[lettersize,journal]{IEEEtran}
\usepackage{amsmath,amsfonts}
\usepackage{algorithmic}
\usepackage{algorithm}
\usepackage{array}
\usepackage[caption=false,font=normalsize,labelfont=sf,textfont=sf]{subfig}
\usepackage{textcomp}
\usepackage{stfloats}
\usepackage{url}
\usepackage{verbatim}
\usepackage{graphicx}
\usepackage{cite}
\begin{document}

\title{Transport Efficiency for Networks Obtained by Vertex Merging Operation}

%\author{IEEE Publication Technology,~\IEEEmembership{Staff,~IEEE,}
        % <-this % stops a space
\author{Zhenhua Yuan, Junhao Peng, and Long Gao
       \IEEEcompsocitemizethanks{\IEEEcompsocthanksitem Zhenhua Yuan, Junhao Peng, and Long Gao are with School of Mathematics and Information Science, Guangzhou University, Guangzhou 510006, China \protect
(E-mail: 1112315004@e.gzhu.edu.cn, pengjh@gzhu.edu.cn, gaolong@gzhu.edu.cn).}

\thanks{This paper was supported by the Basic and Applied Basic Research Foundation of Guangdong Province (Grant No. 2024A1515011700). (Corresponding author:
Junhao Peng.)}% <-this % stops a space
\thanks{Manuscript received XXXX; revised XXXX.}}

%\markboth{IEEE/ACM TRANSACTIONS ON NETWORKING,~Vol.~XX, No.~XX, December~2024}%


\maketitle

\begin{abstract}
Transport is an important function of networks. Studying transport efficiency sheds light on the dynamic processes occurring within various underlying structures and offers a wide range of applications. To construct networks with different transport efficiencies, we focus on the networks obtained by vertex merging operation, which involves connecting multiple graphs through a single node. In this paper, we examine unbiased random walks on these networks and analyze their first-passage properties, including the mean first-passage time (MFPT), the mean trapping time (MTT), and the global-mean first-passage time (GFPT), which characterizes the transport (search) efficiency within the networks. We rigorously derive close-form solutions for these quantities. Results show that all these quantities are governed by the first-passage properties of the constituent components. Additionally, we propose a general method for optimizing the transport (search) efficiency by selecting a suitable node and adjusting the growth of the number of nodes in the subgraphs. We validate our findings using lollipop and barbell graphs. Our results indicate that for an arbitrary GFPT scaling exponent $\alpha \in [1, 3]$, we can construct a network with GFPT scales with the network size $N$ as $\text{GFPT} \sim N^{\alpha}$ through vertex merging operation. These conclusions provide valuable insights for designing and optimizing network structures.
%Transport is an important function of networks. Studying transport efficiency sheds light on the dynamic processes occurring within various underlying structures and offers a wide range of applications. To construct networks with different transport efficiencies, we focus on the networks obtained by vertex merging operation, which involves connecting multiple graphs through a single node. In this paper, we examine unbiased random walks on these networks and analyze their first-passage properties, including the mean first-passage time (MFPT), the mean trapping time (MTT), and the global-mean first-passage time (GFPT), which characterizes the transport (search) efficiency within the networks. We rigorously derive close-form solutions for these quantities. Results show that all these quantities are governed by the first-passage properties of the constituent components. Additionally, we propose a general method for controlling and optimizing the transport (search) efficiency by selecting a suitable node and adjusting the growth of the number of nodes in the subgraphs. We validate our findings using lollipop and barbell graphs. Our results indicate that, for an arbitrary GFPT scaling exponent $\alpha \in [1, 3]$, we can construct a network with GFPT scales with the network size $N$ as $\text{GFPT} \sim N^{\alpha}$ through vertex merging operation. These conclusions provide valuable insights for designing and optimizing network structures.
\end{abstract}


\section{Introduction}
\IEEEPARstart{T}{ransport} of information, energy, and mass is a vital function in many networked systems~\cite{bird-2002-transport, alexandrov-2021-transport, plawsky-2020-transport, alexandrov-2022-transport}. Studying transport problems across various networks helps to characterize and elucidate numerous dynamic processes within these systems and has a wide range of applications in disciplines such as biology, sociology, computer science, and physics~\cite{song-2019-transport, gallos-2007-scaling, hu-2013-adaptation}. Examples include information delivery in mobile ad-hoc networks~\cite{Neely-2005-IEEE-IT, Zhang-2015-IEEE-PDS}, disease transmission or rumor spreading in social networks~\cite{nekovee-2007-theory, yu-2021-modeling, chen-2022-information, Pastor-2015-Epidemic, zhu-2024-defending}, intracellular transport of proteins and other molecular products~\cite{rothman-1994-mechanisms, Bressloff-2013-Stochastic}, energy transfer and animal foraging in food webs~\cite{viswanathan-2011-physics, govenar-2012-energy}, and data collection in sensor networks~\cite{Zheng-2015-IEEE-TPDS, zhang-2018-compressive}.

Global mean first-passage time (GFPT), also known as Kemeny's constant~\cite{kemeny-1969-finite, LO93, Butler2016-book, levene-2002-kemeny, Chen-LMA-2022, Hunter2014-CS}, is a crucial indicator of the information transport efficiency and the stochastic search efficiency of particles within the network. It is defined as the time required for a particle to move randomly to an absorbing node, with both the starting node of the particle and the location of the absorbing node being randomly selected. Previous studies have shown that the GFPT is closely linked to the underlying structure of the network~\cite{ma-2024-determining, QiZhang-2019-IEEE, ben-2000-diffusion, aldous-fill-2014-book, ma-2022-random, ma-2022-structure}. To further explore the impact of network structure on the GFPT, considerable research has focused on analyzing the scaling behavior of the GFPT with network size $N$ in classical symmetry random walks across different network structures. For instance, in extended pseudo-fractal scale-free networks, Apollonian networks, and complete graphs, the GFPT exhibits the smallest scaling behavior, i.e., $\text{GFPT} \sim N$~\cite{Zhang-Lin-2015-PRE, ShengZhang-2019-IEEE-TIT, Bollt-ben-Avraham-2005-NJP}; in recursive small-world trees and Cayley trees, the GFPT scaling is $\text{GFPT} \sim N\log(N)$~\cite{Lin-Zhang-JCP-2013, liu-2013-laplacian}; in comb graphs, $\text{GFPT} \sim N^{\frac{3}{2}}$~\cite{Peng-Elena-PRE-2019}; in tree-like fractals, GFPT grows superlinearly with network size as $\text{GFPT} \sim N^{\alpha}$, where $1<\alpha<2$~\cite{Zhang-Wu-2010-PRE, Lin-Wu-Zhang-2010-PRE, Mafei-2022-method}; in linear chains, GFPT grows quadratically with network size, i.e., $\text{GFPT} \sim N^2$~\cite{aldous-fill-2014-book}; and in barbell graphs, $\text{GFPT} \sim N^3$~\cite{breen-2019-computing, aldous-fill-2014-book, mazo-1982-some}. The exponent in the scaling of GFPT with network size reflects the transport (search) efficiency in a network. The smaller the GFPT scaling exponent, the more efficient the information transport (search) within the network. For a general network with $\text{GFPT} \sim N^\alpha$, its GFPT scaling exponent $\alpha$ is bounded by $1 \leq \alpha \leq 3$~\cite{ShengZhang-2019-IEEE-TIT, Tejedor-Benichou-Voituriez2009-PRE}. According to this measure of transport (search) efficiency, extended pseudo-fractal scale-free networks, Apollonian networks, and complete graphs exhibit the highest transport efficiency, whereas barbell graphs display the lowest transport  efficiency.

Additionally, several studies have focused on enhancing and optimizing transport (search) efficiency by modifying the random walk strategies of particles. Research has shown that incorporating a stochastic resetting strategy into the classical symmetric random walk, where the particles have a certain probability of returning to the starting node at each step, can effectively improve transport efficiency when the resetting rate is appropriately chosen~\cite{Evans_JPA-2011, Wang-2021-CHAOS, Evans-Martin-PRL-2011, chen-2023-optimizing}. Employing a biased random walk strategy by assigning weights to each edge in the network has also demonstrated improved transport efficiency~\cite{gao-2021-optimizing, GaoPeng-2021-Jstat, ZhangPRE-2012a, Peng-Chen-2022-TNSE}. Moreover, maximal-entropy random walks have proven effective in enhancing transport efficiency in certain network configurations~\cite{Peng-Zhang-2014-JCP, niu-2018-maximal}. However, few studies address how to construct networks with diverse transport efficiencies or arbitrary GFPT scaling exponents from existing structures. Developing such networks could help in designing not only efficient wireless networks and radar antennas but also polymers and noncrystalline solids with predetermined diffusion properties and transport efficiencies.

In this paper, we explore the network obtained by vertex merging operation, which involves joining $n$ graphs $\mathcal{G}_1, \mathcal{G}_2, ..., \mathcal{G}_n$ through a single node $C$. Our objective is to construct networks with distinct transport efficiencies and to enable control over the overall transport efficiency. We analyze the first-passage properties of unbiased random walk on this class of networks, focusing on key indicators such as mean first-passage time (MFPT), mean trapping time (MTT), and GFPT. By employing probabilistic methods, we derive analytic solutions for these quantities. Our results demonstrate that all these quantities are determined by the first-passage property of the individual graphs $\mathcal{G}_1, \mathcal{G}_2, ..., \mathcal{G}_n$. Furthermore, we propose adjusting the position of node $C$ and controlling the growth of the total number of nodes in graphs $\mathcal{G}_1, \mathcal{G}_2, ..., \mathcal{G}_n$ to optimize and enhance overall transport efficiency. Lollipop graphs and barbell graphs serve as examples to validate our findings. Finally, we introduce a method for constructing networks with arbitrary GFPT scaling exponent $\alpha \in [1,3]$. The conclusions obtained here contribute to the advancement of network construction and the optimization of network structures.

The remainder of this paper is organized as follows. Sec.\ref{sec:PRELIMINARIES} introduces the vertex merging operation and discusses the quantities analyzed in the transport (search) problem, including the mean first-passage time (MFPT), the mean trapping time (MTT), and the GFPT. Sec.\ref{sec: DIMENSIONALITY} analyzes the MFPT between any starting and absorbing nodes, where the starting and absorbing nodes differ. Furthermore, Sec.\ref{sec:Mean} investigates the MTT for any given absorbing node. Sec.\ref{sec:Kemeny_constant} presents the exact results for the GFPT of the overall structure. Sec.\ref{sec:Optimization} presents a general method for controlling transport (search) efficiency and provides two illustrative examples. Finally, Sec.\ref{sec:Conclusion} summarizes the findings of this paper.

\section{Preliminaries}
\label{sec:PRELIMINARIES}
\subsection{Vertex Merging Operation}
\begin{figure}
\begin{center}
\includegraphics[width=0.45 \textwidth]{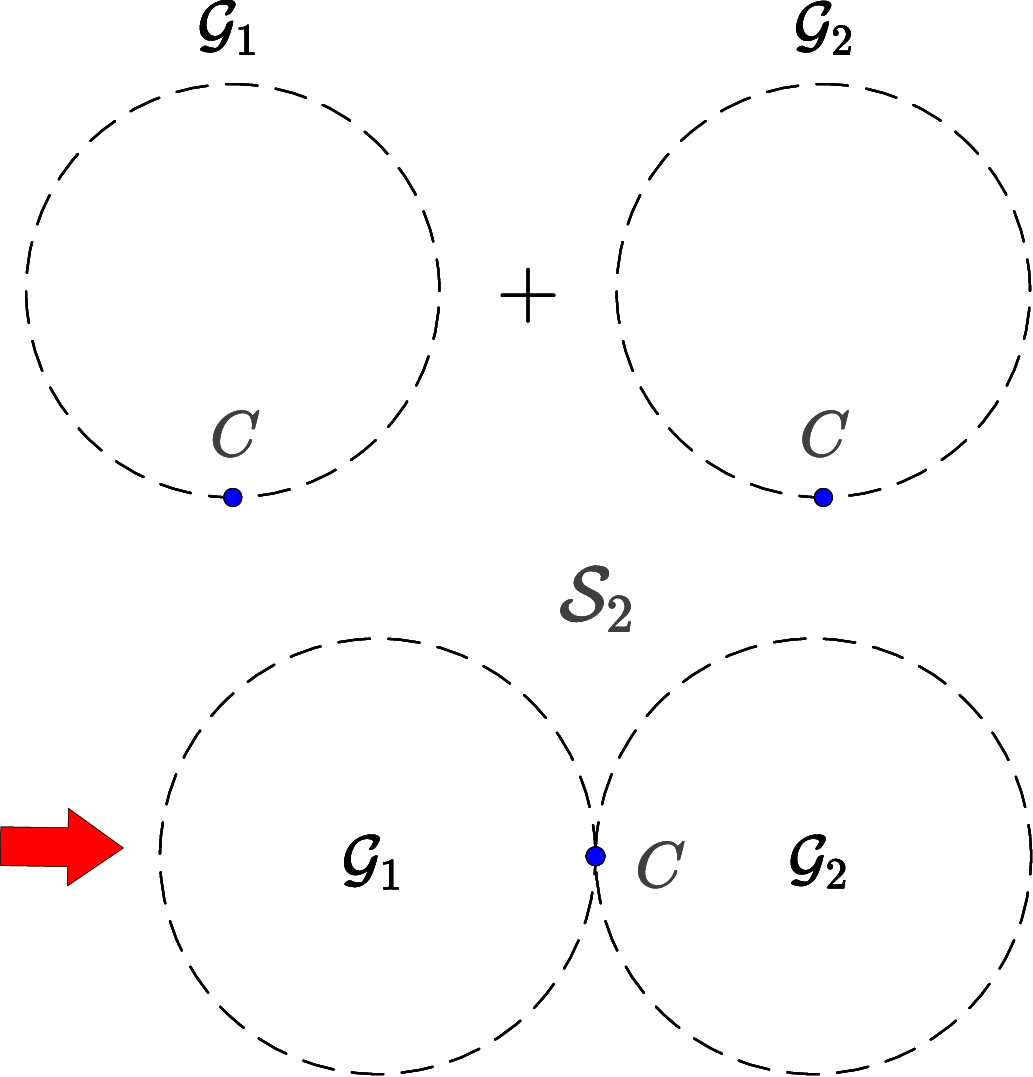}
\caption{A schematic illustration of the vertex merging operation. The blue node $C$ is the hub of the graph $\mathcal{S}_2$. The graphs $\mathcal{G}_1$ and $\mathcal{G}_2$ are joined together through vertex $C$ to form the graph $\mathcal{S}_2=(\mathcal{G}_1, \mathcal{G}_2)$, which is depicted on the right-hand side of the arrow.
}
\label{fig:1}
\end{center}
\end{figure}

To begin, we need to prepare $n$ $(n \geq 2)$ simple connected graphs $\mathcal{G}_1, \mathcal{G}_2, ..., \mathcal{G}_n$. Let $V_{\mathcal{G}_i}$ $(i=1, 2, ..., n)$ be the set of nodes of graph $\mathcal{G}_i$. The vertex merging operation involves joining the graphs $\mathcal{G}_1, \mathcal{G}_2, ..., \mathcal{G}_n$ through designated $n$ vertices $C_i \in V_{\mathcal{G}_i}$ $(i=1, 2, ..., n)$, such that vertices $C_1, C_2, ..., C_n$ coincide. We denote the resulting graph obtained from the vertex merging operation as $\mathcal{S}_n=(\mathcal{G}_1, \mathcal{G}_2, ..., \mathcal{G}_n)$, with the overlapped vertex referred to as $C$, which serves as the ``hub" of $\mathcal{S}_n$. Clearly, the hub $C$ is a cut vertex of graph $\mathcal{S}_n$, and $\mathcal{G}_1, \mathcal{G}_2, ..., \mathcal{G}_n$ are the subgraphs of $\mathcal{S}_n$. For convenience, we will also refer to $C_i$ in subgraph $\mathcal{G}_i$ as node $C$. Fig.~\ref{fig:1} illustrates a schematic diagram of the vertex merging operation using $\mathcal{S}_2=(\mathcal{G}_1, \mathcal{G}_2)$ as an example.

Let $N_i$ be the total number of nodes in subgraph $\mathcal{G}_i$ and $M_i$ be the total number of edges in subgraph $\mathcal{G}_i$. The total number of nodes in graph $\mathcal{S}_n$, denoted by $N^{\mathcal{S}_n}$, is
\begin{equation}\label{NK7}
N^{\mathcal{S}_n}=\sum_{k=1}^n N_k-n+1 \,,
\end{equation}
and the total number of edges in graph $\mathcal{S}_n$, denoted by $M^{\mathcal{S}_n}$, is
\begin{equation}\label{NK14}
M^{\mathcal{S}_n}= \sum_{k=1}^n M_k \,.
\end{equation}
Through vertex merging operation, a wide variety of graphs can be constructed by varying the structure of subgraphs and the position of node $C$ within each subgraph. Additionally, some special graphs can also be obtained through this operation. For example, lollipop graphs can be created by combining a complete graph with a path, and barbell graphs can be formed by joining a lollipop graph with a complete graph~\cite{aldous-fill-2014-book}.

%First, we need to prepare $n$ connected graphs $\mathcal{G}_1, $\mathcal{G}_2, ..., $\mathcal{G}_n$. Let $V_i$ $(i=1, 2)$ be the set of nodes of the graph $\mathcal{G}_i$. The vertex merging operation is to join $\mathcal{G}_1$ and $\mathcal{G}_2$ through the given two vertices $C_1 \in V_1$ and $C_2 \in V_2$, such that vertex $C_1$ overlaps with vertex $C_2$. We denote the overlapped vertex as $C$ and the graph obtained by the vertex merging operation of graphs $\mathcal{G}_1$ and $\mathcal{G}_2$ as $\mathcal{S}_2=(\mathcal{G}_1, \mathcal{G}_2)$. Clearly, vertex $C$ is a cut vertex of graph $\mathcal{S}_2$. For convenience, we also refer to $C_i$ $(i=1, 2)$ in the graph $\mathcal{G}_i$ as node $C$. Fig.~\ref{fig:1} illustrates a schematic diagram of the vertex merging operation.

%In particular, if node $C$ is still a cut vertex of subgraph $\mathcal{G}_1$, then we can continue the decomposition of subgraph $\mathcal{G}_1$ as above. Therefore, we write the graph $\mathcal{S}_n=(\mathcal{G}_1, \mathcal{G}_2, ..., \mathcal{G}_n)$, indicating that the graph $\mathcal{S}_n$ can be decomposed into $\mathcal{G}_1$, $\mathcal{G}_2 ,..., \mathcal{G}_n$ of $n$ subgraphs, where the node $C$ is not a cut vertex of any subgraphs $\mathcal{G}_i$ $(i=1, 2, ..., n)$. In this way, a complex graph can be decomposed into several simple subgraphs to analyze its first-passage properties.

%%%%%%%%%%%%%%%%%%%%%%%%%%%%%%%%%%%%%%%%%%%%%%%%%%%%%%%%%
% Figure  01
%%%%%%%%%%%%%%%%%%%%%%%%%%%%%%%%%%%%%%%%%%%%%%%%%%%%%%%%%%

\subsection{First Passage Time}
Consider an unbiased random walk of a particle on graph $\mathcal{S}_n$. At each time step, the particle moves randomly and uniformly to one of the neighbors of its current node until it becomes trapped at the absorbing node. The average time required for the particle to move from the starting node $A$ to the absorbing node $B$ is known as the mean first-passage time (MFPT), referred to as $\left\langle T_{A \rightarrow B}^{\mathcal{S}_n} \right\rangle$, which is closely related to the positions of the starting node and absorbing node~\cite{Book-Redner-2007, Condamin-Benichou-Tejedor-2007-Nature, ma-2024-structural}.

If the starting node of the particle is an absorbing node, the MFPT is recovered by the mean first-return time (MFRT), referred to as $\left\langle T_{A \rightarrow A}^{\mathcal{S}_n} \right\rangle$. Let $d_A^{\mathcal{S}_n}$ represent the degree of node $A$ in graph $\mathcal{S}_n$, and let $V_{\mathcal{S}_n}$ be the set of nodes of graph $\mathcal{S}_n$. Then, the MFRT of node $A$ on the graph $\mathcal{S}_n$ can be expressed as~\cite{Noh-Rieger-2004-PRL, condamin-2007-random, aldous-fill-2014-book}
\begin{equation}\label{BNK1}
\left\langle T_{A \rightarrow A}^{\mathcal{S}_n} \right\rangle = \frac{1}{\pi_A^{\mathcal{S}_n}} \,,
\end{equation}
where $\pi_A^{\mathcal{S}_n}$ is the stationary occupation probability of node $A$ in $\mathcal{S}_n$, with
\begin{equation}\label{BNK2}
\pi_A^{\mathcal{S}_n} = \frac{d_A^{\mathcal{S}_n}}{\sum_{B \in V_{\mathcal{S}_n}}d_B^{\mathcal{S}_n}} \,.
\end{equation}

To comprehensively measure the dynamic properties of absorbing node $B$ in graph $\mathcal{S}_n$, one considers the starting node of the particle across all possible positions to obtain the MFPT, known as the mean trapping time (MTT) of node $B$, referred to as $\left\langle T_{B}^{\mathcal{S}_n} \right\rangle$, which is defined as~\cite{Hwang-Lee-Kahng-2012-PRL, Gao-Peng-2022-PRE}
\begin{equation}\label{BNK3}
\left\langle T_{B}^{\mathcal{S}_n} \right\rangle = \sum_{A \in V_{\mathcal{S}_n}} \pi_{A}^{\mathcal{S}_n} \left\langle T_{A\rightarrow B}^{\mathcal{S}_n}  \right\rangle   \,.
\end{equation}
The MTT is closely related to the location of node $B$ and serves as an important indicator of node centrality and the information transport efficiency of a node.

In addition, the MFPT obtained by taking over all the possible positions for the starting and absorbing nodes is the global mean first-passage time (GFPT) of the network $\mathcal{S}_n$, also known as Kemeny's constant, referred to as $\left\langle T^{\mathcal{S}_n} \right\rangle$, which is defined as~\cite{PENG-Sandev-Kocarev2020-CNSNS, Catral-2010-JSC, PitmanTang-Bernoulli-2018, xia-2024-efficient}
\begin{equation}\label{BNK4}
\left\langle T^{\mathcal{S}_n} \right\rangle = \sum_{B \in V_{\mathcal{S}_n}} \pi_{B}^{\mathcal{S}_n} \sum_{A \in V_{\mathcal{S}_n}} \pi_{A}^{\mathcal{S}_n} \left\langle T_{A \rightarrow B}^{\mathcal{S}_n} \right\rangle   \,.
\end{equation}
The GFPT depends on the underlying structure of the graph $\mathcal{S}_n$ and serves as an important indicator of the transport (search) efficiency within the network.

\subsection{Notations}
Table \ref{table:1} lists the notations frequently used in the paper and their descriptions.
\begin{table}[!t]
\renewcommand\arraystretch{1.3}
\begin{center}
\centering
\caption{Notation Explanations.}
\label{table:1}
\setlength{\tabcolsep}{1mm}{
\resizebox{\linewidth}{!}{
\begin{tabular}{c c}
\hline
  Notation & Description\\
\hline
  $n$  &  The number of subgraphs comprising the graph $\mathcal{S}_n$ \\
  $C$  &  The hub of graph $\mathcal{S}_n$ \\
  $A^{\mathcal{G}_i}$ &  The node $A$ in subgraph $\mathcal{G}_i$\\
  $\overline{\mathcal{G}_{i}}$  &The induced subgraph obtained by removing all nodes  in \\
   &subgraph $\mathcal{G}_{i}$ (except node $C$) and their connected edges in $\mathcal{S}_n$\\
  $N^{\mathcal{G}}$ &  Total number of nodes in graph $\mathcal{G}$\\
  $M^{\mathcal{G}}$ &  Total number of edges in graph $\mathcal{G}$\\
  $N_{i}$ &  Total number of nodes in subgraph $\mathcal{G}_i$\\
  $M_{i}$ &  Total number of edges in subgraph $\mathcal{G}_i$\\
  $V_{\mathcal{G}}$ &  The set of nodes in graph $\mathcal{G}$\\
  $d_i^{\mathcal{G}}$ &   The degree of node $i$ in graph $\mathcal{G}$\\
  $\pi_i^{\mathcal{G}}$ &   Stationary occupation probability of node $i$ in graph $\mathcal{G}$\\
  $T_{i \rightarrow j}^{\mathcal{G}}$ &  The first-passage time from node $i$ to $j$ in graph $\mathcal{G}$ \\
  $\left\langle T_{i \rightarrow j}^{\mathcal{G}}  \right\rangle$ &  The MFPT from node $i$ to $j$ in graph $\mathcal{G}$ \\
  $\left\langle T_{i}^{\mathcal{G}} \right\rangle $ &  The MTT of node $i$ in graph $\mathcal{G}$ \\
  $\left\langle T^{\mathcal{G}} \right\rangle $ &  The GFPT of the graph $\mathcal{G}$ \\ \hline
\end{tabular}}}
\end{center}
\end{table}

\section{Mean first-passage time}
\label{sec: DIMENSIONALITY}
This section focuses on the mean first-passage time (MFPT) from starting node $A$ to the absorbing node $B$, where $A\neq B$. We will analyze the MFPT between any two nodes in the network obtained by the vertex merging operation. The discussion is organized into three subsections Sec.\ref{sec:base_base}-Sec.\ref{sec:base_fiber} based on the location of the starting node and absorbing node.

\subsection{One of the Starting Node and the Absorbing Node Is the Hub}
\label{sec:base_base}
Here, we analyze the MFPT between hub $C$ and an arbitrary node that is not hub $C$ in graph $\mathcal{S}_n$. We will apply probability theory to compute their MFPT within $\mathcal{S}_n$.

We begin by analyzing the MFPT from an arbitrary node $A$ to hub $C$ in $\mathcal{S}_n$, where $A \neq C$. Given the construction of $\mathcal{S}_n$, there exists a positive integer $i$ $(1 \leq i \leq n)$ such that $A\in V_{\mathcal{G}_i}$. Let $A^{\mathcal{G}_i}$ denote the node $A$ in subgraph $\mathcal{G}_i$. Since the only intersection between different subgraphs is hub $C$, clearly, the MFPT from $A$ to hub $C$ in graph $\mathcal{S}_n$, denoted by $\left\langle T_{A^{\mathcal{G}_i} \rightarrow C}^{\mathcal{S}_n}  \right\rangle$, is
\begin{equation}\label{NK1}
\left\langle T_{A^{\mathcal{G}_i} \rightarrow C}^{\mathcal{S}_n} \right\rangle = \left\langle T_{A \rightarrow C}^{\mathcal{G}_i} \right\rangle \,.
\end{equation}

Next, we consider the MFPT from hub $C$ to an arbitrary node $B$ in graph $\mathcal{S}_n=(\mathcal{G}_1, \mathcal{G}_2, ..., \mathcal{G}_n)$, where $B \neq C$. Similarly, there exists a positive integer $i$ $(1 \leq i \leq n)$ such that $B\in V_{\mathcal{G}_i}$. Let $D$ be the event of the particle's first step moving to the subgraph $\mathcal{G}_i$, and let event $\bar{D}$ be its complement. The probabilities of events $D$ and $\bar{D}$ occurring are $P(D)=\frac{d_{C}^{\mathcal{G}_i}}{d_C^{\mathcal{S}_n}}$ and $P(\bar{D})=1-\frac{d_{C}^{\mathcal{G}_i}}{d_C^{\mathcal{S}_n}}$, respectively, where $d_k^{\mathcal{G}}$ is the degree of node $k$ in graph $\mathcal{G}$. Let $T_{C \rightarrow B^{\mathcal{G}_i}}^{\mathcal{S}_n}$ be the random variable representing the first-passage time from $C$ to $B$ in graph $\mathcal{S}_n$, and let $E(T_{C \rightarrow B^{\mathcal{G}_i}}^{\mathcal{S}_n}|D)$ stand for the conditional expectation of the random variable $T_{C \rightarrow B^{\mathcal{G}_i}}^{\mathcal{S}_n}$ given that the event $D$ occurs. By applying the law of total expectation~\cite{chung-2000-course}, the MFPT from hub $C$ to $B$ in graph $\mathcal{S}_n$, denoted by $\left\langle T_{C \rightarrow B^{\mathcal{G}_i}}^{\mathcal{S}_n} \right\rangle$, can be expressed as
\begin{alignat}{1}\label{NK2}
\left\langle T_{C \rightarrow B^{\mathcal{G}_i}}^{\mathcal{S}_n} \right\rangle = P(D) \times & E(T_{C \rightarrow B^{\mathcal{G}_i}}^{\mathcal{S}_n}|D) \nonumber\\
&+ P(\bar{D}) \times E(T_{C \rightarrow B^{\mathcal{G}_i}}^{\mathcal{S}_n}|\bar{D}) \,.
\end{alignat}

When event $\bar{D}$ occurs, since the only intersection between different subgraphs is hub $C$, the particle must return to $C$ before visiting node $B$. Let $\overline{\mathcal{G}_i}$ denote the induced subgraph formed by removing all nodes in subgraph $\mathcal{G}_i$ (except node $C$) along with their connecting edges in graph $\mathcal{S}_n$~\cite{west-2001-introduction}, and let $\left\langle T_{C \rightarrow C}^{\overline{\mathcal{G}_i}} \right\rangle$ be the mean first-return time of node $C$ in induced subgraph $\overline{\mathcal{G}_i}$. Then, we have
\begin{equation}\label{NK3}
E(T_{C \rightarrow B^{\mathcal{G}_i}}^{\mathcal{S}_n}|\bar{D}) = \left\langle T_{C \rightarrow C}^{\overline{\mathcal{G}_i}} \right\rangle + \left\langle T_{C \rightarrow B^{\mathcal{G}_i}} ^{\mathcal{S}_n} \right\rangle \,,
\end{equation}
where
\begin{equation}
\left\langle T_{C \rightarrow C}^{\overline{\mathcal{G}_i}} \right\rangle=\frac{2(\sum_{k=1}^n M_k-M_i)}{d_{C}^{\mathcal{S}_n} - d_{C}^{\mathcal{G}_i}} \,,
\end{equation}
with $M_k$ being the total number of edges of subgraph $\mathcal{G}_k$, as derived from Eqs.~(\ref{BNK1}) and (\ref{BNK2}).

When event $D$ occurs, let $P(T_{C \rightarrow C}^{\mathcal{G}_i} > T_{C \rightarrow B}^{\mathcal{G}_i})$ be the probability that the particle starting from node $C$ visits $B$ before returning to $C$ in subgraph $\mathcal{G}_i$, and $\pi_{C}^{\mathcal{G}_i}$ be the stationary occupation probability of node $C$ in subgraph $\mathcal{G}_i$. By applying the law of total expectation again, we have
\begin{alignat}{1}\label{NK4}
&E(T_{C \rightarrow B^{\mathcal{G}_i}}^{\mathcal{S}_n}|D)= P(T_{C \rightarrow C}^{\mathcal{G}_i} > T_{C \rightarrow B}^{\mathcal{G}_i})
\nonumber\\
&\times E(T_{C \rightarrow B}^{\mathcal{G}_i}|T_{C \rightarrow C}^{\mathcal{G}_i} > T_{C \rightarrow B}^{\mathcal{G}_i})+P(T_{C \rightarrow C}^{\mathcal{G}_i} < T_{C \rightarrow B}^{\mathcal{G}_i})
\nonumber\\
&\times  \bigg[E(T_{C \rightarrow C}^{\mathcal{G}_i}|T_{C \rightarrow C}^{\mathcal{G}_i} < T_{C \rightarrow B}^{\mathcal{G}_i})+\left\langle T_{C \rightarrow B^{\mathcal{G}_i}}^{\mathcal{S}_n} \right\rangle \bigg] \,,
\end{alignat}
where~\cite{aldous-fill-2014-book} $$P(T_{C \rightarrow C}^{\mathcal{G}_i} > T_{C \rightarrow B}^{\mathcal{G}_i})=\frac{1}{\pi_{C}^{\mathcal{G}_i} \bigg[ \left\langle T_{C \rightarrow B}^{\mathcal{G}_i} \right\rangle +\left\langle T_{B \rightarrow C}^{\mathcal{G}_i} \right\rangle \bigg]}$$
and the stationary occupation probability $\pi_{C}^{\mathcal{G}_i}$ is given by Eq.~(\ref{BNK2}). For subgraph $\mathcal{G}_i$, the MFPT from node $C$ to $B$, denoted by $\left\langle T_{C \rightarrow B}^{\mathcal{G}_i} \right\rangle$, is expressed as
\begin{alignat}{1}\label{NK5}
&\left\langle T_{C \rightarrow B}^{\mathcal{G}_i} \right\rangle = P(T_{C \rightarrow C}^{\mathcal{G}_i} > T_{C \rightarrow B}^{\mathcal{G}_i})
\nonumber\\
&\times E(T_{C \rightarrow B}^{\mathcal{G}_i}|T_{C \rightarrow C}^{\mathcal{G}_i} > T_{C \rightarrow B}^{\mathcal{G}_i})+P(T_{C \rightarrow C}^{\mathcal{G}_i} < T_{C \rightarrow B}^{\mathcal{G}_i})
\nonumber\\
&\times  \bigg[E(T_{C \rightarrow C}^{\mathcal{G}_i}|T_{C \rightarrow C}^{\mathcal{G}_i} < T_{C \rightarrow B}^{\mathcal{G}_i})+\left\langle T_{C \rightarrow B}^{\mathcal{G}_i} \right\rangle \bigg] \,.
\end{alignat}
Substituting Eq.~(\ref{NK5}) into Eq.~(\ref{NK4}), we have
\begin{alignat}{1}\label{NK6}
E(T_{C \rightarrow B^{\mathcal{G}_i}}^{\mathcal{S}_n}|D) =& P(T_{C \rightarrow C}^{\mathcal{G}_i} < T_{C \rightarrow B}^{\mathcal{G}_i}) \times \left\langle T_{C \rightarrow B^{\mathcal{G}_i}}^{\mathcal{S}_n} \right\rangle
\nonumber\\
&+ P(T_{C \rightarrow C}^{\mathcal{G}_i} > T_{C \rightarrow B}^{\mathcal{G}_i}) \times \left\langle T_{C \rightarrow B}^{\mathcal{G}_i} \right\rangle  \,.
\end{alignat}
Inserting Eqs.~(\ref{NK3}) and (\ref{NK6}) into Eq.~(\ref{NK2}), we eventually obtain
\begin{alignat}{1}\label{NK8}
\left\langle T_{C \rightarrow B^{\mathcal{G}_i}}^{\mathcal{S}_n} \right\rangle =&\frac{\sum_{k=1}^{n} M_k}{M_i} \times \left\langle T_{C \rightarrow B}^{\mathcal{G}_i} \right\rangle
\nonumber\\
&+ \bigg(\frac{\sum_{k=1}^{n} M_k}{M_i}-1 \bigg)\times \left\langle T_{B \rightarrow C}^{\mathcal{G}_i} \right\rangle \,.
\end{alignat}
That is to say, in graph $\mathcal{S}_n$, the MFPT from the hub $C$ to an arbitrary absorbing node $B$ in subgraph $\mathcal{G}_i$ is determined by the commute time between $C$ and $B$ in subgraph $\mathcal{G}_i$.

\subsection{The Starting Node and the Absorbing Node (Neither of Which Is the Hub) Are on the Same Subgraph}
\label{sec:fiber_base}
In this subsection, we consider the MFPT from a starting node $A$ and an absorbing node $B$ in graph $\mathcal{S}_n$, where both nodes are not the hub $C$ and are located within the same subgraph.

By the construction of graph $\mathcal{S}_n$, we can assume that both nodes $A$ and $B$ are within subgraph $\mathcal{G}_i$, where $i$ is a positive integer satisfying $1 \leq i \leq n$. Let $P(T_{A \rightarrow B}^{\mathcal{G}_i} > T_{A \rightarrow C}^{\mathcal{G}_i})$ denote the probability that the particle starting from node $A$ visits node $C$ before reaching node $B$ in subgraph $\mathcal{G}_i$. According to the law of total expectation, the MFPT from $A$ to $B$ in graph $\mathcal{S}_n$, denoted by $\left\langle T_{A^{\mathcal{G}_i} \rightarrow B^{\mathcal{G}_i}}^{\mathcal{S}_n} \right\rangle$, can be expressed as
\begin{alignat}{1}\label{NK9}
&\left\langle T_{A^{\mathcal{G}_i} \rightarrow B^{\mathcal{G}_i}}^{\mathcal{S}_n} \right\rangle= P(T_{A \rightarrow B}^{\mathcal{G}_i} < T_{A \rightarrow C}^{\mathcal{G}_i})
\nonumber\\
&\times E(T_{A \rightarrow B}^{\mathcal{G}_i}|T_{A \rightarrow B}^{\mathcal{G}_i} < T_{A \rightarrow C}^{\mathcal{G}_i})+P(T_{A \rightarrow B}^{\mathcal{G}_i} > T_{A \rightarrow C}^{\mathcal{G}_i})
\nonumber\\
&\times  \bigg[E(T_{A \rightarrow C}^{\mathcal{G}_i}|T_{A \rightarrow B}^{\mathcal{G}_i} > T_{A \rightarrow C}^{\mathcal{G}_i})+\left\langle T_{C \rightarrow B^{\mathcal{G}_i}} ^{\mathcal{S}_n} \right\rangle \bigg] \,,
\end{alignat}
where~\cite{aldous-fill-2014-book} $$P(T_{A \rightarrow B}^{\mathcal{G}_i} > T_{A \rightarrow C}^{\mathcal{G}_i})=\frac{\left\langle T_{A \rightarrow B}^{\mathcal{G}_i} \right\rangle+ \left\langle T_{B \rightarrow C}^{\mathcal{G}_i} \right\rangle - \left\langle T_{A \rightarrow C}^{\mathcal{G}_i} \right\rangle}{\left\langle T_{C \rightarrow B}^{\mathcal{G}_i} \right\rangle +\left\langle T_{B \rightarrow C}^{\mathcal{G}_i} \right\rangle} \,.$$
Similarly, for subgraph $\mathcal{G}_i$, we have
\begin{alignat}{1}\label{NK10}
&\left\langle T_{A \rightarrow B}^{\mathcal{G}_i} \right\rangle= P(T_{A \rightarrow B}^{\mathcal{G}_i} < T_{A \rightarrow C}^{\mathcal{G}_i})
\nonumber\\
&\times E(T_{A \rightarrow B}^{\mathcal{G}_i}|T_{A \rightarrow B}^{\mathcal{G}_i} < T_{A \rightarrow C}^{\mathcal{G}_i})+P(T_{A \rightarrow B}^{\mathcal{G}_i} > T_{A \rightarrow C}^{\mathcal{G}_i})
\nonumber\\
&\times  \bigg[E( T_{A \rightarrow C}^{\mathcal{G}_i}|T_{A \rightarrow B}^{\mathcal{G}_i} > T_{A \rightarrow C}^{\mathcal{G}_i})+\left\langle  T_{C \rightarrow B}^{\mathcal{G}_i} \right\rangle \bigg] \,.
\end{alignat}
Merging Eqs.~(\ref{NK9}) and (\ref{NK10}), we obtain
\begin{alignat}{1}\label{NK11}
\left\langle T_{A^{\mathcal{G}_i} \rightarrow B^{\mathcal{G}_i}}^{\mathcal{S}_n} \right\rangle =&\left\langle T_{A \rightarrow B}^{\mathcal{G}_i} \right\rangle+ P(T_{A \rightarrow B}^{\mathcal{G}_i} > T_{A \rightarrow C}^{\mathcal{G}_i})
\nonumber\\
&\times \bigg[ \left\langle T_{C \rightarrow B^{\mathcal{G}_i}}^{\mathcal{S}_n} \right\rangle - \left\langle T_{C \rightarrow B}^{\mathcal{G}_i} \right\rangle \bigg] \,.
\end{alignat}
Recalling the expression for $\left\langle T_{C \rightarrow B^{\mathcal{G}_i}}^{\mathcal{S}_n} \right\rangle $ in Eq.~(\ref{NK8}), we ultimately obtain
\begin{alignat}{1}\label{NK12}
\left\langle T_{A^{\mathcal{G}_i} \rightarrow B^{\mathcal{G}_i}}^{\mathcal{S}_n} \right\rangle& =\frac{\sum_{k=1}^{n} M_k}{M_i} \times \left\langle T_{A \rightarrow B}^{\mathcal{G}_i} \right\rangle
\nonumber\\
+& \bigg( \frac{\sum_{k=1}^{n} M_k}{M_i}-1\bigg)\times \bigg[ \left\langle T_{B \rightarrow C}^{\mathcal{G}_i} \right\rangle- \left\langle T_{A \rightarrow C}^{\mathcal{G}_i} \right\rangle \bigg]\,.
\end{alignat}
That is, in graph $\mathcal{S}_n$, the MFPT between any two nodes in subgraph $\mathcal{G}_i$ is related to their MFPT in subgraph $\mathcal{G}_i$, the MFPT from $A$ to $C$ in subgraph $\mathcal{G}_i$, and the MFPT from $B$ to $C$ in subgraph $\mathcal{G}_i$.

\subsection{The Starting Node and the Absorbing Node (Neither of Which Is the Hub) Are on Different Subgraphs}
\label{sec:base_fiber}
This subsection focuses on the MFPT between any two nodes in the graph $\mathcal{S}_n$ that belong to different subgraphs. Without loss of generality, we assume that the starting node $A$ is in subgraph $\mathcal{G}_i$ and the absorbing node $B$ is in subgraph $\mathcal{G}_j$ $(j \neq i)$, and that neither $A$ nor $B$ is the hub~\footnote{A similar conclusion holds if the starting node $A$ is in subgraph $\mathcal{G}_j$ and the absorbing node $B$ is in subgraph $\mathcal{G}_i$.}.

Since nodes $A$ and $B$ are in different subgraphs, the particle must visit hub $C$ before reaching node $B$. Thus, the MFPT from node $A$ to $B$ in graph $\mathcal{S}_n$, denoted by $\left\langle T_{A^{\mathcal{G}_i}  \rightarrow B^{\mathcal{G}_j}}^{\mathcal{S}_n} \right\rangle$, is
\begin{equation}\label{NK13}
\left\langle T_{A^{\mathcal{G}_i}  \rightarrow B^{\mathcal{G}_j}}^{\mathcal{S}_n}\right\rangle =\left\langle T_{A  \rightarrow C}^{\mathcal{G}_i} \right\rangle + \left\langle T_{C \rightarrow B^{\mathcal{G}_j}}^{\mathcal{S}_n} \right\rangle \,.
\end{equation}
Recalling Eq.~(\ref{NK8}), we obtain
\begin{alignat}{1}\label{NK15}
\left\langle T_{A^{\mathcal{G}_i}  \rightarrow B^{\mathcal{G}_j}}^{\mathcal{S}_n}\right\rangle =&\left\langle T_{A  \rightarrow C}^{\mathcal{G}_i} \right\rangle + \frac{\sum_{k=1}^{n} M_k}{M_j} \times \left\langle T_{C \rightarrow B}^{\mathcal{G}_j} \right\rangle
\nonumber\\
&+ \bigg( \frac{\sum_{k=1}^{n} M_k}{M_j} -1 \bigg)\times \left\langle T_{B \rightarrow C}^{\mathcal{G}_j} \right\rangle \,.
\end{alignat}
In other words, in graph $\mathcal{S}_n$, the MFPT from a starting node $A$ in subgraph $\mathcal{G}_i$ to an absorbing node $B$ in subgraph $\mathcal{G}_j$ is closely related to the MFPT from $A$ to $C$ in subgraph $\mathcal{G}_i$, and the commute time between $B$ and $C$ in subgraph $\mathcal{G}_j$.

In summary, this section rigorously derives the MFPT between any two nodes in graph $\mathcal{S}_n$. We demonstrate that the MFPT is characterized by the first-passage properties within the subgraphs, which helps to predict the efficiency of information transport between any two nodes in graph $\mathcal{S}_n$.

\section{Mean trapping time} \label{sec:Mean}
Here, we turn to analyze the mean trapping time (MTT) of an arbitrary node in the network obtained by vertex merging operation. Our goal is to establish a relation between the MTT of the nodes in graph $\mathcal{S}_n$ and the MTT in the corresponding subgraph. We will begin by examining the case of $\mathcal{S}_2=(\mathcal{G}_1, \mathcal{G}_2)$ and subsequently generalize our findings to the case of $\mathcal{S}_n=(\mathcal{G}_1, \mathcal{G}_2, ..., \mathcal{G}_n)$. Our results will show that the MTT for the nodes in $\mathcal{S}_n$ can be expressed as a function of the MTT of the nodes in the subgraphs.

\subsection{MTT for the Hub}
In this subsection, we assume the hub $C$ is an absorbing node, and we focus on evaluating the MTT of hub $C$ in the graph $\mathcal{S}_n$.

%Let $\left\langle T_{i \rightarrow j}^{\mathcal{S}_n} \right\rangle$ be the MFPT from nodes $i$ to $j$ in $\mathcal{S}_n$ and $V_{\mathcal{G}}$ be the set of nodes of graph $\mathcal{G}$. The MTT for the absorbing node $j$, denoted by $\left\langle T_j^{\mathcal{S}_n}\right\rangle$, is defined as
%\begin{equation}\label{NK16}
%\left\langle T_{j}^{\mathcal{S}_n} \right\rangle = \sum_{i \in V_{\mathcal{S}_n}} \pi_{i}^{\mathcal{S}_n} \left\langle T_{i\rightarrow j}^{\mathcal{S}_n}  \right\rangle   \,.
%\end{equation}

For the graph $\mathcal{S}_2=(\mathcal{G}_1, \mathcal{G}_2)$, based on Eq.~(\ref{NK1}) and the definition in Eq.~(\ref{BNK3}), the MTT of the hub $C$ in graph $\mathcal{S}_2$, denoted by $\left\langle T_{C}^{\mathcal{S}_2} \right\rangle$, can be expressed as
\begin{alignat}{1}\label{NK17}
\left\langle T_{C}^{\mathcal{S}_2} \right\rangle =& \pi_C^{\mathcal{S}_2} \left\langle T_{C \rightarrow C}^{\mathcal{S}_2} \right\rangle +\sum_{\substack{A \in V_{\mathcal{G}_1} \\ A\neq C}} \bigg[ \frac{d_A^{\mathcal{G}_1}}{2(M_1+M_2)} \left\langle T_{A \rightarrow C}^{\mathcal{G}_1} \right\rangle \bigg]
\nonumber\\
&+  \sum_{\substack{ A \in V_{\mathcal{G}_2} \\ A\neq C}} \bigg( \frac{d_A^{\mathcal{G}_2}}{2(M_1+M_2)}\times  \left\langle T_{A \rightarrow C}^{\mathcal{G}_2} \right\rangle  \bigg) \,.
\end{alignat}
For the subgraph $\mathcal{G}_i$ $(i=1,2)$, the MTT of node $C$, denoted by $\left\langle T_{C}^{\mathcal{G}_i} \right\rangle$, is expressed as
\begin{equation}\label{NK18}
\left\langle T_{C}^{\mathcal{G}_i} \right\rangle = 1+ \sum_{\substack{A \in V_{\mathcal{G}_i}\\ A\neq C}} \bigg( \frac{d_A^{\mathcal{G}_i}}{2M_i}\times  \left\langle T_{A \rightarrow C}^{\mathcal{G}_i} \right\rangle \bigg)  \,.
\end{equation}
Substituting Eq.~(\ref{NK18}) into Eq.~(\ref{NK17}), we obtain
\begin{equation}\label{NK19}
\left\langle  T_{C}^{\mathcal{S}_2} \right\rangle = \frac{M_1}{M_1+M_2} \times  \left\langle T_{C}^{\mathcal{G}_1} \right\rangle + \frac{M_2}{M_1+M_2} \times \left\langle T_{C}^{\mathcal{G}_2} \right\rangle\,.
\end{equation}

For the graph $\mathcal{S}_n=(\mathcal{G}_1, \mathcal{G}_2, ..., \mathcal{G}_n)$, it can be viewed as being constructed by joining $\mathcal{G}_n$ and $\mathcal{S}_{n-1}$ through the node $C$, as illustrated in Fig.~\ref{fig:1}. Consequently, following Eqs.~(\ref{NK14}) and (\ref{NK19}), the MTT of hub $C$ in graph $\mathcal{S}_n$, $\left\langle  T_{C}^{\mathcal{S}_n} \right\rangle$, is expressed as
\begin{equation}\label{NK50}
\left\langle  T_{C}^{\mathcal{S}_n} \right\rangle =\frac{M_n}{\sum_{k=1}^n M_k}  \times \left\langle T_{C}^{\mathcal{G}_n} \right\rangle +\frac{\sum_{k=1}^{n-1} M_k}{\sum_{k=1}^n M_k}  \times \left\langle T_{C}^{\mathcal{S}_{n-1}} \right\rangle  \,.
\end{equation}
Eventually, we recursively obtain
\begin{equation}\label{NK20}
\left\langle  T_{C}^{\mathcal{S}_n} \right\rangle =\frac{1}{\sum_{k=1}^n M_k}  \times \sum_{i=1}^n  \bigg(M_i \times \left\langle T_{C}^{\mathcal{G}_i} \right\rangle \bigg)\,.
\end{equation}
Note that the right-hand side of Eq.~(\ref{NK20}) represents a weighted average of the MTT of node $C$ across different subgraphs $\mathcal{G}_i$ $(i=1,2,...,n)$, weighted by the total number of edges in each subgraph. Therefore, the MTT of hub $C$ in graph $\mathcal{S}_n$ is actually the average of the MTT of $C$ across the various subgraphs.

Furthermore, Eq.~(\ref{NK20}) suggests that, given $n$ subgraphs, selecting the node with the highest node transport efficiency among them, i.e., the node with the lowest MTT, can significantly enhance the transport efficiency of hub $C$ in graph $\mathcal{S}_n$. This approach not only improves the transport efficiency of the hub but also contributes to optimizing the overall network structure.

\subsection{MTT for an Arbitrary Absorbing Node Except the Hub}
In this subsection, we focus on the MTT of an arbitrary absorbing node $B$, which is distinct from the hub $C$. Without loss of generality, we assume that absorbing node $B$ is located in subgraph $\mathcal{G}_i$, where $i$ is a positive integer satisfying $1 \leq i \leq n$.

We first examine the case of graph $\mathcal{S}_2=(\mathcal{G}_1, \mathcal{G}_2)$. Recall that $\overline{\mathcal{G}_i}$ denotes the induced subgraph obtained by removing all nodes in subgraph $\mathcal{G}_i$ (except for node $C$) along with their connecting edges in graph $\mathcal{S}_n$. Utilizing the definition provided in Eq.~(\ref{BNK3}) and combining Eqs.~(\ref{NK8}), (\ref{NK12}), and (\ref{NK15}), we ultimately derive the MTT of node $B$ in graph $\mathcal{S}_2$, denoted by $\left\langle T_{B^{\mathcal{G}_i}}^{\mathcal{S}_2} \right\rangle$, as
\begin{alignat}{1}\label{NK21}
& \left\langle T_{B^{\mathcal{G}_i}}^{\mathcal{S}_2} \right\rangle =\sum_{\substack{A \in V_{\mathcal{G}_i} \\ A\neq C}} \bigg( \pi^{\mathcal{S}_2}_A \times \left\langle T_{A^{\mathcal{G}_i}\rightarrow B^{\mathcal{G}_i}}^{\mathcal{S}_2}  \right\rangle\bigg)
\nonumber\\
&+ \sum_{\substack{A \in V_{\overline{\mathcal{G}_i}} \\ A\neq C}} \bigg(\pi^{\mathcal{S}_2}_A \times \left\langle T_{A^{\overline{\mathcal{G}_i}}\rightarrow B^{\mathcal{G}_i}}^{\mathcal{S}_2} \right\rangle\bigg) +\pi^{\mathcal{S}_2}_C \times \left\langle T_{C \rightarrow B^{\mathcal{G}_i}}^{\mathcal{S}_2} \right\rangle
\nonumber\\
=& \sum_{\substack{A \in V_{\mathcal{G}_i} \\ A\neq C}} \bigg\{\frac{d_A^{\mathcal{G}_i}}{2(M_1+M_2)} \times \bigg[\frac{M_1+M_2}{M_i}  \times  \left\langle T_{A\rightarrow B}^{\mathcal{G}_i}  \right\rangle
\nonumber\\
&+\bigg(\frac{M_1 +M_2}{M_i}-1\bigg) \times \bigg(\left\langle T_{B\rightarrow C}^{\mathcal{G}_i}  \right\rangle - \left\langle T_{A\rightarrow C}^{\mathcal{G}_i}  \right\rangle \bigg) \bigg] \bigg\}
\nonumber\\
&+\sum_{\substack{A \in V_{\overline{\mathcal{G}_i}} \\ A\neq C}} \bigg\{\frac{d_A^{\overline{\mathcal{G}_i}}}{2(M_1+M_2)}\times \bigg[ \left\langle T_{A\rightarrow C}^{\overline{\mathcal{G}_i}} \right\rangle + \frac{M_1+M_2}{M_i}
\nonumber\\
&\times \left\langle T_{C \rightarrow B}^{\mathcal{G}_i}  \right\rangle +\bigg(\frac{M_1+M_2}{M_i}-1 \bigg) \times \left\langle T_{B \rightarrow C}^{\mathcal{G}_i}  \right\rangle \bigg]\bigg\}
\nonumber\\
&+\frac{d_{C}^{\mathcal{G}_1}+d_{C}^{\mathcal{G}_2}}{2(M_1+M_2)} \times \bigg[ \frac{M_1+M_2}{M_i} \times \left\langle T_{C \rightarrow B}^{\mathcal{G}_i} \right\rangle
\nonumber\\
&+\bigg( \frac{M_1+M_2}{M_i}-1 \bigg)\times \left\langle T_{B \rightarrow C}^{\mathcal{G}_i} \right\rangle   \bigg]
\nonumber\\
=& \bigg(1- \frac{M_i}{M_1+M_2} \bigg)\times \bigg( \left\langle T_{C}^{\overline{\mathcal{G}_i}}  \right\rangle -\left\langle T_{C}^{\mathcal{G}_i}  \right\rangle \bigg) +\left\langle T_{B}^{\mathcal{G}_i} \right\rangle
\nonumber\\
&+\bigg(\frac{M_1+M_2}{M_i}-1\bigg)\times \bigg( \left\langle T_{C \rightarrow B}^{\mathcal{G}_i}  \right\rangle+\left\langle T_{B \rightarrow C}^{\mathcal{G}_i}  \right\rangle \bigg) \,,
\end{alignat}
where $\left\langle T_{B}^{\mathcal{G}_i} \right\rangle$ is the MTT of node $B$ in subgraph $\mathcal{G}_i$. In particular, when $i = 2$, the subgraph $\overline{\mathcal{G}_2}$ corresponds to the subgraph $\mathcal{G}_1$, and Eq.~(\ref{NK21}) can be rewritten as
\begin{alignat}{1}\label{BNK21}
\left\langle T_{B^{\mathcal{G}_2}}^{\mathcal{S}_2} \right\rangle =& \frac{M_1}{M_1+M_2}\times \bigg( \left\langle T_{C}^{\mathcal{G}_1}  \right\rangle -\left\langle T_{C}^{\mathcal{G}_2}  \right\rangle \bigg)+\left\langle T_{B}^{\mathcal{G}_2} \right\rangle
\nonumber\\
&+\frac{M_1}{M_2}\times \bigg( \left\langle T_{C \rightarrow B}^{\mathcal{G}_2}  \right\rangle + \left\langle T_{B \rightarrow C}^{\mathcal{G}_2}  \right\rangle \bigg) \,.
\end{alignat}

In the following, we consider the MTT for absorbing node $B \in V_{\mathcal{G}_i}$ within graph $\mathcal{S}_n=(\mathcal{G}_1, \mathcal{G}_2, ..., \mathcal{G}_n)$, where $n \geq 2$. The graph ${\mathcal{S}_n}$ can be viewed as being constructed by joining subgraphs $\mathcal{G}_i$ and $\overline{\mathcal{G}_{i}}$ through node $C$. According to Eq.~(\ref{NK21}), the MTT of node $B$ in graph $\mathcal{S}_n$ can be expressed in terms of the first-passage properties of subgraph $\mathcal{G}_i$ and induced subgraph $\overline{\mathcal{G}_{i}}$. Consequently, we have
\begin{alignat}{1}\label{NK22}
\left\langle T_{B^{\mathcal{G}_i}}^{\mathcal{S}_n} \right\rangle  = &\bigg(\frac{\sum_{k=1}^{n} M_k}{M_i}-1 \bigg) \times \bigg( \left\langle T_{C \rightarrow B}^{\mathcal{G}_i}  \right\rangle +\left\langle T_{B \rightarrow C}^{\mathcal{G}_i}  \right\rangle \bigg)
\nonumber\\
&+\bigg(1-\frac{M_i}{\sum_{k=1}^n M_k} \bigg) \times \bigg( \left\langle T_{C}^{\overline{\mathcal{G}_{i}}}  \right\rangle -\left\langle T_{C}^{\mathcal{G}_i}  \right\rangle \bigg)
\nonumber\\
&+\left\langle T_{B}^{\mathcal{G}_i} \right\rangle\,.
\end{alignat}
From the construction of the induced subgraph $\overline{\mathcal{G}_i}$, it follows that Eq.~(\ref{NK20}) remains valid. Thus, we have
\begin{equation}\label{NK23}
\left\langle  T_{C}^{\overline{\mathcal{G}_{i}}} \right\rangle =\frac{ \sum_{k=1}^{n}  \bigg(M_k \times \left\langle T_{C}^{\mathcal{G}_k} \right\rangle \bigg) -M_i \times \left\langle T_{C}^{\mathcal{G}_i} \right\rangle}{\sum_{k=1}^{n} M_k -M_i} \,.
\end{equation}
Inserting Eq.~(\ref{NK23}) into Eq.~(\ref{NK22}) and using the relation $\left\langle T_{B}^{\mathcal{G}_i}  \right\rangle - \left\langle T_C^{\mathcal{G}_i}  \right\rangle= \left\langle T_{C \rightarrow B}^{\mathcal{G}_i}  \right\rangle- \left\langle T_{B \rightarrow C}^{\mathcal{G}_i}  \right\rangle$~\cite{aldous-fill-2014-book}, we finally get
\begin{alignat}{1}\label{NK24}
\left\langle T_{B^{\mathcal{G}_i}}^{\mathcal{S}_n} \right\rangle  =&\frac{\sum_{k=1}^n \bigg(M_k \times \left\langle T_{C}^{\mathcal{G}_k} \right\rangle \bigg) }{\sum_{k=1}^n M_k} +\frac{\sum_{k=1}^n M_k }{M_i} \left\langle T_{C\rightarrow B}^{\mathcal{G}_i} \right\rangle
\nonumber\\
&+\bigg(\frac{\sum_{k=1}^{n} M_k }{M_i} -2 \bigg) \times \left\langle T_{B \rightarrow C}^{\mathcal{G}_i} \right\rangle \,.
\end{alignat}
This implies that in graph $\mathcal{S}_n$, the MTT of an absorbing node $B$ in the subgraph $\mathcal{G}_i$ is determined by the weighted average of the MTT of node $C$ across each subgraph $\mathcal{G}_k$ $(k=1,2,...,n)$, along with the commute time between nodes $B$ and $C$ in subgraph $\mathcal{G}_i$.

Overall, we find that the MTT of any given absorbing node in graph $\mathcal{S}_n$ is expressed in terms of the first-passage properties (i.e., MFPT and MTT) on the subgraphs. MTT serves as a crucial measure of the transport efficiency of nodes within a network. The conclusions obtained here contribute to predicting the transport efficiency of nodes in graph $\mathcal{S}_n$, offering a more detailed understanding of how the structure of the subgraphs influences overall transport dynamics.

\section{Global mean first-passage time} \label{sec:Kemeny_constant}
In this section, we focus on the global mean first-passage time (GFPT) of the network obtained through the vertex merging operation, exploiting the connection between the GFPT of the graph $\mathcal{S}_n$ and the MTT and GFPT of its subgraphs. We begin by demonstrating the exact result for graph $\mathcal{S}_2=(\mathcal{G}_1, \mathcal{G}_2)$, and then employ mathematical induction to generalize our findings to the case of $\mathcal{S}_n=(\mathcal{G}_1, \mathcal{G}_2, ..., \mathcal{G}_n)$.

For the graph $\mathcal{S}_2=(\mathcal{G}_1, \mathcal{G}_2)$, by inserting the expressions from Eqs.~(\ref{NK19}) and (\ref{NK21}) into Eq.~(\ref{BNK4}), we obtain the GFPT of graph $\mathcal{S}_n$, denoted by $\left\langle T^{\mathcal{S}_2} \right\rangle$, as
\begin{alignat}{1}\label{NK26}
&\left\langle T^{\mathcal{S}_2} \right\rangle = \pi^{\mathcal{S}_2}_C \times \left\langle T_{C}^{\mathcal{S}_2} \right\rangle  +\sum_{\substack{A \in V_{\mathcal{G}_1} \\ A\neq C}} \bigg( \pi^{\mathcal{S}_2}_A \times \left\langle T_{A}^{\mathcal{S}_2}  \right\rangle\bigg)
\nonumber\\
&+ \sum_{\substack{A \in V_{\mathcal{G}_2} \\ A\neq C}} \bigg(\pi^{\mathcal{S}_2}_A \times \left\langle T_{A}^{\mathcal{S}_2}  \right\rangle\bigg)
\nonumber\\
=&\frac{d_{C}^{\mathcal{G}_1}+d_{C}^{\mathcal{G}_2}}{2(M_1+M_2)} \times \bigg( \frac{M_1}{M_1+M_2}  \left\langle T_{C}^{\mathcal{G}_1} \right\rangle + \frac{M_2}{M_1+M_2}  \left\langle T_{C}^{\mathcal{G}_2} \right\rangle \bigg)
\nonumber\\
&+\sum_{\substack{A \in V_{\mathcal{G}_1} \\ A\neq C}} \bigg\{\frac{d_A^{\mathcal{G}_1}}{2(M_1+M_2)}\times \bigg[\frac{M_2}{M_1+M_2}\times \bigg( \left\langle T_{C}^{\mathcal{G}_2}  \right\rangle -
\nonumber\\
&\left\langle T_{C}^{\mathcal{G}_1}  \right\rangle \bigg) +\frac{M_2}{M_1}\times \bigg( \left\langle T_{C \rightarrow A}^{\mathcal{G}_1}  \right\rangle +\left\langle T_{A \rightarrow C}^{\mathcal{G}_1}  \right\rangle \bigg) +\left\langle T_{A}^{\mathcal{G}_1} \right\rangle \bigg] \bigg\}
\nonumber\\
&+\sum_{\substack{A \in V_{\mathcal{G}_2} \\ A\neq C}} \bigg\{\frac{d_A^{\mathcal{G}_2}}{2(M_1+M_2)}\times \bigg[\frac{M_1}{M_1+M_2}\times \bigg( \left\langle T_{C}^{\mathcal{G}_1}  \right\rangle -
\nonumber\\
&\left\langle T_{C}^{\mathcal{G}_2}  \right\rangle \bigg) +\frac{M_1}{M_2}\times \bigg( \left\langle T_{C \rightarrow A}^{\mathcal{G}_2}  \right\rangle +\left\langle T_{A \rightarrow C}^{\mathcal{G}_2}  \right\rangle \bigg) +\left\langle T_{A}^{\mathcal{G}_2} \right\rangle \bigg] \bigg\}
\nonumber\\
=&\left\langle T^{\mathcal{G}_1} \right\rangle +\left\langle T^{\mathcal{G}_2} \right\rangle + \frac{M_2}{M_1+M_2}\times \left\langle T_{C}^{\mathcal{G}_1}  \right\rangle \nonumber\\
&+\frac{M_1}{M_1+M_2}\times \left\langle T_{C}^{\mathcal{G}_2}  \right\rangle-2 \,,
\end{alignat}
where $\left\langle T^{\mathcal{G}_i} \right\rangle$ is the GFPT of subgraph $\mathcal{G}_i$.

For the graph $\mathcal{S}_n=(\mathcal{G}_1, \mathcal{G}_2, ..., \mathcal{G}_n)$, where $n \geq 2$, it can be viewed as being constructed by joining $\mathcal{G}_n$ and $\mathcal{S}_{n-1}$ through the node $C$. Consequently, following Eq.~(\ref{NK26}), we obtain the recursive formula for the GFPT of graph $\mathcal{S}_{n}$:
\begin{alignat}{1}\label{ANK4}
\left\langle T^{\mathcal{S}_{n}} \right\rangle = \left\langle T^{\mathcal{G}_n} \right\rangle &+\left\langle T^{\mathcal{S}_{n-1}} \right\rangle + \frac{\sum_{i=1}^{n-1} M_i}{\sum_{i=1}^n M_i}\times \left\langle T_{C}^{\mathcal{G}_n}  \right\rangle
\nonumber\\ &+\frac{M_n}{\sum_{i=1}^n M_i}\times \left\langle T_{C}^{\mathcal{S}_{n-1}}  \right\rangle-2   \,.
\end{alignat}
Recalling Eq.~(\ref{NK20}) and using Eq.~(\ref{ANK4}) recursively, we finally obtain
\begin{alignat}{1}\label{NK27}
\left\langle T^{\mathcal{S}_n} \right\rangle =& \sum_{i=1}^n \left\langle T^{\mathcal{G}_i} \right\rangle + \sum_{j=1}^n \bigg[ \bigg( 1-\frac{M_j}{\sum_{i=1}^n M_i}  \bigg) \times \left\langle T_C^{\mathcal{G}_j} \right\rangle \bigg]
\nonumber\\ &-2(n-1)\,.
\end{alignat}
Eq.~(\ref{NK27}) indicates that the transport efficiency of graph $\mathcal{S}_n$ is primarily determined by the GFPT of each subgraph $\mathcal{G}_i$ $(i=1,2,.... ,n)$ and the selection of node $C$ within these subgraphs.

GFPT is an important measure of information transport efficiency and stochastic search efficiency in networks. In Sec.\ref{sec:Optimization}, we discuss the application of Eq.~(\ref{NK27}) in optimizing the network structure and constructing the network. Our findings reveal that for any predefined GFPT scaling exponent $\alpha \in [1, 3]$, networks with $\text{GFPT} \sim (N)^{\alpha}$ can be constructed by selecting suitable subgraphs and performing the vertex merging operation, where $N$ is the network size.

Moreover, Eq.~(\ref{NK27}) effectively predicts the transport efficiency of large-scale graphs $\mathcal{S}$ with numerous cut vertices, significantly reducing the computational demands and time consumption. For such large-scale graphs, simulating the GFPT can be memory-intensive, as it requires storing extensive node information.  Additionally, the simulation process involves evaluating all possible configurations of the starting node and absorbing node within the graph $\mathcal{S}$, leading to considerable time expenditure. Accurately computing the GFPT for large-scale graphs is challenging, particularly due to the need to compute the pseudoinverse of the large-scale Laplacian matrix of $\mathcal{S}$~\cite{Lin-Wu-Zhang-2010-PRE, banerjee-1973-generalized}. However, Eq.~(\ref{NK27}) allows us to first decompose the graph $\mathcal{S}$ into multiple simpler subgraphs using the cut vertices, reversing the process illustrated in Fig.~\ref{fig:1}. Subsequently, the GFPT of graph $\mathcal{S}$ can be computed through the first-passage properties of these subgraphs, including their MTT and GFPT.

\section{Control and optimization of network transport efficiency} \label{sec:Optimization}
In this section, we will first present a general method for controlling and optimizing the transport efficiency of networks obtained through vertex merging operation, and then illustrate our conclusions using lollipop graphs and barbell graphs as examples. By adjusting the position of node $C$ and the growth of the number of nodes in the subgraph, we will demonstrate that, for any arbitrary GFPT scaling exponent $\alpha \in [1, 3]$, it is possible to create a network with GFPT scaling with the network size $N$ as $\text{GFPT} \sim N^{\alpha}$.

\subsection{General Method}
\label{sec:Generalized_methods}
Here, we use the expression from Eq.~(\ref{NK26}) to present a generalized method for controlling network transport efficiency.

Let $N_i$ denote the total number of nodes in graph $\mathcal{G}_i$ $(i=1, 2)$. The network size of graph $\mathcal{S}_2$, denoted by $N^{\mathcal{S}_2}$, is given by  $N^{\mathcal{S}_2}= N_1+N_2-1$. We assume that the scaling of GPFT in the subgraph $\mathcal{G}_i$ $(i=1, 2)$ is $\left\langle T^{\mathcal{G}_i} \right\rangle \sim (N_i)^{\alpha_i}$, where $\alpha_1 \geq \alpha_2 $. To control the scaling behavior of the GFPT in graph $\mathcal{S}_2$, we introduce the parameter $\lambda$ such that $N_1=(N_2)^\lambda$. In the following, we will examine the effect of different values of $\lambda$ on the transport efficiency of $\mathcal{S}_2$ across various cases.

%%%%%%%%%%%%%%%%%%%%%%%%%%%%%%%%%%%%%%%%%%%%%%%%%%%%%%%%%
% Figure  01
%%%%%%%%%%%%%%%%%%%%%%%%%%%%%%%%%%%%%%%%%%%%%%%%%%%%%%%%%%
\begin{figure}
\begin{center}
\includegraphics[width=0.48 \textwidth]{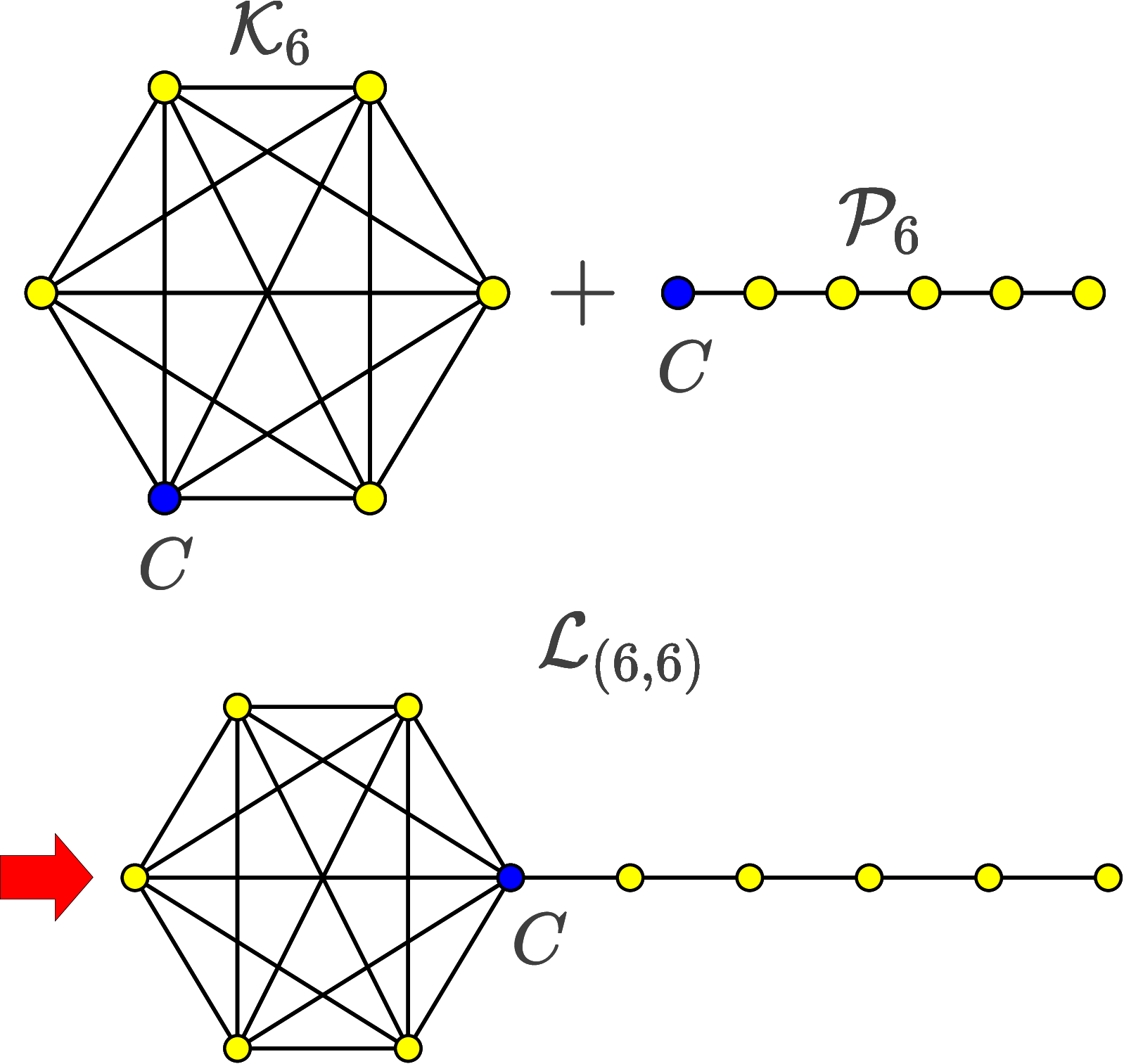}
\caption{A schematic illustration of the construction for the lollipop graph $\mathcal{L}_{(6,6)}$. The lollipop graph $\mathcal{L}_{(6,6)}$, shown on the right-hand side of the arrow, is formed by connecting a complete graph $\mathcal{K}_{6}$ and a path $\mathcal{P}_{6}$ through node $C$.
}
\label{fig:2}
\end{center}
\end{figure}

First, we consider the case where the MTT of the hub $C$ does not affect the scaling of GFPT in graph $\mathcal{S}_2$, i.e., $\left\langle T^{\mathcal{S}_2} \right\rangle \sim \left\langle T^{\mathcal{G}_1} \right\rangle + \left\langle T^{\mathcal{G}_2} \right\rangle$.

\noindent \emph{Case 1}: If $\lambda<\frac{\alpha_2}{\alpha_1}$, then we have $N^{\mathcal{S}_2} \sim N_2$ and $\left\langle T^{\mathcal{S}_2} \right\rangle  \sim (N_2)^{\lambda\alpha_1}+(N_2)^{\alpha_2} \sim (N^{\mathcal{S}_2})^{\alpha_2}$. In this case, the transport efficiency of graph $\mathcal{S}_2$ is comparable to that of $\mathcal{G}_2$.

\noindent \emph{Case 2}: If $\frac{\alpha_2}{\alpha_1}\leq \lambda \leq 1$, then $N^{\mathcal{S}_2} \sim N_2$ and $\left\langle T^{\mathcal{S}_2} \right\rangle  \sim (N_2)^{\lambda\alpha_1} \sim (N^{\mathcal{S}_2})^{\lambda\alpha_1}$, where $\alpha_2 \leq \lambda\alpha_1 \leq \alpha_1$. In this case, the transport efficiency of graph $\mathcal{S}_2$ lies between that of the subgraphs $\mathcal{G}_1$ and $\mathcal{G}_2$.

\noindent \emph{Case 3}: If $\lambda >1$, then we have $N^{\mathcal{S}_2} \sim (N_2)^{\lambda}$ and $\left\langle T^{\mathcal{S}_2} \right\rangle  \sim (N_2)^{\lambda\alpha_1}\sim (N^{\mathcal{S}_2})^{\alpha_1}$. The transport efficiency of graph $\mathcal{S}_2$ is comparable to that of $\mathcal{G}_1$.

\noindent To summarize, when $\alpha_1 \geq \alpha_2$, we have
\begin{equation}
\left\langle T^{\mathcal{S}_2} \right\rangle  \sim \left\{
\begin{array}{ll}
  (N^{\mathcal{S}_2})^{\alpha_2},   & {\text{if~}} \lambda<\frac{\alpha_2}{\alpha_1}, \\ \\
  (N^{\mathcal{S}_2})^{\lambda\alpha_1},   &{\text{if~}} \frac{\alpha_2}{\alpha_1}\leq \lambda \leq 1,\\ \\
  (N^{\mathcal{S}_2})^{\alpha_1},  & {\text{otherwise.}}
   \end{array}
  \right.
\label{NK34}
\end{equation}
That is, by selecting different subgraphs and adjusting the parameter $\lambda$, we can create networks with varying transport efficiencies\textemdash or, more precisely, with different GFPT scaling exponents $\alpha \in [\alpha_2, \alpha_1]$.

Next, we consider the case where the MTT of hub $C$ influences the scaling of GFPT in graph $\mathcal{S}_2$. To construct networks with diverse transport efficiencies, we select the node with the largest MTT in subgraphs $\mathcal{G}_1$ and $\mathcal{G}_2$ as the node $C$. We assume it is located in subgraph $\mathcal{G}_1$, and its MTT scales with the network sizes as $\left\langle T^{\mathcal{G}_1}_C  \right\rangle \sim (N_1)^{\beta}$, where $\beta \geq \alpha_1$. Let the total number of edges in subgraph $\mathcal{G}_i$ scale with the network size such that $M_i \sim (N_i)^{\gamma_i}$, where $i=1, 2$. According to Eq.~(\ref{NK26}), for $N_2 \to \infty$, we obtain
\begin{alignat}{1}\label{NK35}
\left\langle T^{\mathcal{S}_2} \right\rangle &\sim   \left\langle T^{\mathcal{G}_1} \right\rangle + \left\langle T^{\mathcal{G}_2} \right\rangle + \frac{M_2}{M_1+M_2} \left\langle T_C^{\mathcal{G}_1} \right\rangle
\nonumber\\ & \sim (N_2)^{\lambda \alpha_1}+(N_2)^{\alpha_2}+\frac{(N_2)^{\gamma_2}}{(N_2)^{\lambda \gamma_1}+(N_2)^{\gamma_2}} (N_2)^{\lambda\beta}\,.
\end{alignat}
If $\lambda \leq {\text{\emph{min}}}(\frac{\gamma_2}{\gamma_1}, 1)$, then we have $N^{\mathcal{S}_2} \sim N_2$, and Eq.~(\ref{NK35}) can rewritten as
\begin{alignat}{1}\label{NK36}
\left\langle T^{\mathcal{S}_2} \right\rangle  \sim (N^{\mathcal{S}_2})^{\eta}\,,
\end{alignat}
where $\alpha_2 \leq \eta \leq \lambda\beta$. This indicates that by adjusting parameter $\lambda$, we can construct a network with an arbitrary GFPT scaling exponent $\alpha \in [\alpha_2, \lambda \beta]$. The selection of the node $C$ is equally important in building networks with varying transport efficiencies.

\subsection{Examples}
\label{sec:Examples}
This subsection demonstrates our conclusions using lollipop graphs and barbell graphs as examples. We will accurately calculate the GFPT of the lollipop graph and subsequently present a specific method to control the transport efficiency in both lollipop graphs and barbell graphs.

The lollipop graph $\mathcal{L}_{(n_1,n_2)}$ is formed by joining a complete graph $\mathcal{K}_{n_1}$ of size $n_1$ on the endpoint $C$ of a path $\mathcal{P}_{n_2}$ of size $n_2$~\cite{aldous-fill-2014-book}. Fig.~\ref{fig:2} depicts the construction of the lollipop graph $\mathcal{L}_{(6,6)}$. One can see that the total number of nodes in the lollipop graph is $N^{\mathcal{L}_{(n_1,n_2)}}=n_1+n_2-1$, while the total number of edges of subgraphs $\mathcal{K}_{n_1}$ and $\mathcal{P}_{n_2}$ are $M_1=\frac{n_1(n_1-1)}{2}$ and $M_2=n_2-1$, respectively.

Previous research~\cite{aldous-fill-2014-book} has shown that the GFPT and the MTT of any node $C\in V_{\mathcal{K}_{n_1}}$ in a complete graph $\mathcal{K}_{n_1}$ are given by
\begin{equation}\label{NK28}
\left\langle T^{\mathcal{K}_{n_1}} \right\rangle = \left\langle T_{C}^{\mathcal{K}_{n_1}} \right\rangle =\frac{(n_1-1)^2}{n_1} +1\,.
\end{equation}
Additionally, the MTT of endpoint $C$ in a path $\mathcal{P}_{n_2}$ is
\begin{equation}\label{NK29}
\left\langle T_C^{\mathcal{P}_{n_2}} \right\rangle = \frac{2(n_2-1)^2}{3}+\frac{5}{6}\,,
\end{equation}
and the GFPT is
\begin{equation}\label{NK30}
\left\langle T^{\mathcal{P}_{n_2}} \right\rangle = \frac{(n_2-1)^2}{3}+\frac{7}{6}  \,.
\end{equation}
By inserting Eqs.~(\ref{NK28}), (\ref{NK29}), and (\ref{NK30}) into Eq.~(\ref{NK26}), we obtain the GPFT of the lollipop graph, denoted by $\left\langle T^{\mathcal{L}_{(n_1,n_2)}} \right\rangle$, as
\begin{alignat}{1}\label{NK31}
&\left\langle T^{\mathcal{L}_{(n_1,n_2)}} \right\rangle = \frac{3(n_1)^3(n_2)^2-6(n_1)^3n_2-3(n_1)^2(n_2)^2}{6n_1n_2+3(n_1)^3-3(n_1)^2-6n_1}
\nonumber\\
&+ \frac{18(n_1)^2n_2+2n_1(n_2)^3-6n_1(n_2)^2-11n_1n_2}{6n_1n_2+3(n_1)^3-3(n_1)^2-6n_1}
\nonumber\\
&+ \frac{3(n_1)^4-3(n_1)^3-9(n_1)^2+12n_1+12n_2-12}{6n_1n_2+3(n_1)^3-3(n_1)^2-6n_1}\,.
\end{alignat}
For small-scale networks, the GFPT can be accurately calculated using MATLAB~\cite{Peng-Stanley-2018-JSTAT}. We have compared the results of Eq.~(\ref{NK31}) with the numerical results obtained from MATLAB, and they are in complete agreement.

%%%%%%%%%%%%%%%%%%%%%%%%%%%%%%%%%%%%%%%%%%%%%%%%%%%%%%%%%
% Figure  01
%%%%%%%%%%%%%%%%%%%%%%%%%%%%%%%%%%%%%%%%%%%%%%%%%%%%%%%%%%
\begin{figure}
\begin{center}
\includegraphics[width=0.46 \textwidth]{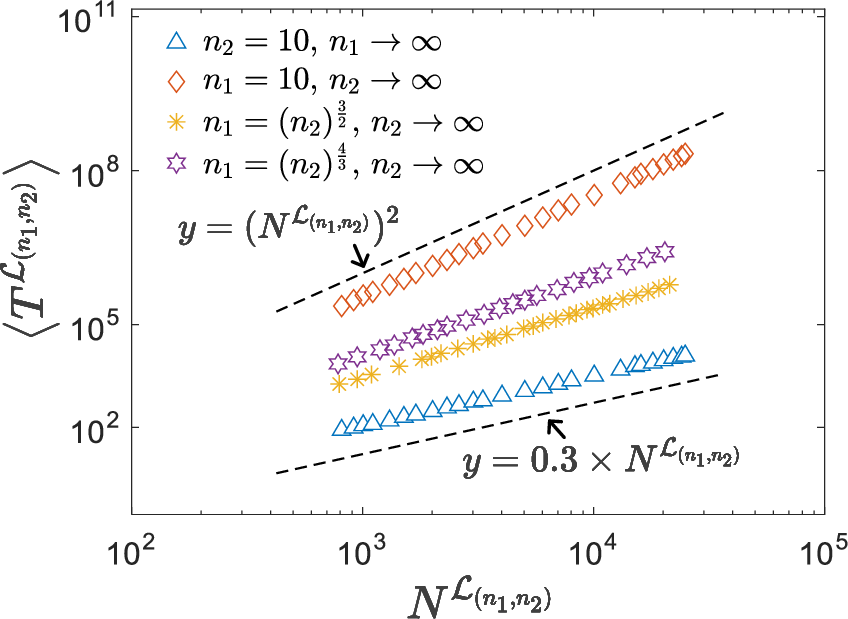}
\caption{Double logarithmic plots of $\left\langle T^{\mathcal{L}_{(n_1,n_2)}} \right\rangle$ versus $N^{\mathcal{L}_{(n_1,n_2)}}$ $(N^{\mathcal{L}_{(n_1,n_2)}}=n_1+n_2-1)$ in the lollipop graphs for different choices of $n_1$ and $n_2$. The dashed lines serve as references. The GFPT of the lollipop graphs exhibit different scaling behaviors under different growth of the number of nodes in the subgraphs.
}
\label{fig:3}
\end{center}
\end{figure}

%For networks of small sizes, the probability generating function of the MFPT can be readily computed using the symbolic Math Toolbox of MATLAB, which leads to the MFPT between any two nodes and the GFPT of the network~\cite{Peng-Stanley-2018-JSTAT}. We have verified that these results are consistent with Eq.~(\ref{NK31}).

In the following, we demonstrate how to control the transport efficiency of a lollipop graph. To construct a network with the GFPT scales as $\left\langle T^{\mathcal{L}_{(n_1,n_2)}} \right\rangle \sim (N^{\mathcal{L}_{(n_1,n_2)}})^\beta$, where $1 \leq \beta \leq 2$, we can set $$n_1=(n_2)^\lambda \,, $$ with $1 \leq \lambda \leq 2$. As $n_2$ approaches infinity, $$N^{\mathcal{L}_{(n_1,n_2)}}=n_1+n_2-1 \sim (n_2)^\lambda \,.$$ According to Eq.~(\ref{NK26}), for $n_2 \rightarrow \infty$, we find $$\left\langle T^{\mathcal{L}_{(n_1,n_2)}} \right\rangle \sim \left\langle T^{\mathcal{K}_{n_1}} \right\rangle+\left\langle T^{\mathcal{P}_{n_2}} \right\rangle \sim (N^{\mathcal{L}_{(n_1,n_2)}})^{\frac{2}{\lambda}}.$$Therefore, the predefined GFPT scaling exponent $\beta$ can be achieved by taking $\lambda= \frac{2}{\beta}$. Fig.~\ref{fig:3} demonstrates that the GFPT of the lollipop graph exhibits different scaling behaviors depending on the growth of the number of nodes in the subgraph. In particular, for the case where $n_1$ is fixed and $n_2 \rightarrow \infty$, we have $ \left\langle T^{\mathcal{L}_{(n_1,n_2)}} \right\rangle \sim (N^{\mathcal{L}_{(n_1,n_2)}})^2$, while for the case where $n_1=(n_2)^{\frac{4}{3}}$ and $n_2 \rightarrow \infty $, we have $\left\langle T^{\mathcal{L}_{(n_1,n_2)}} \right\rangle \sim (N^{\mathcal{L}_{(n_1,n_2)}})^{\frac{3}{2}}$, which aligns with our previous conclusions.

\begin{figure}
\begin{center}
\includegraphics[width=0.48 \textwidth]{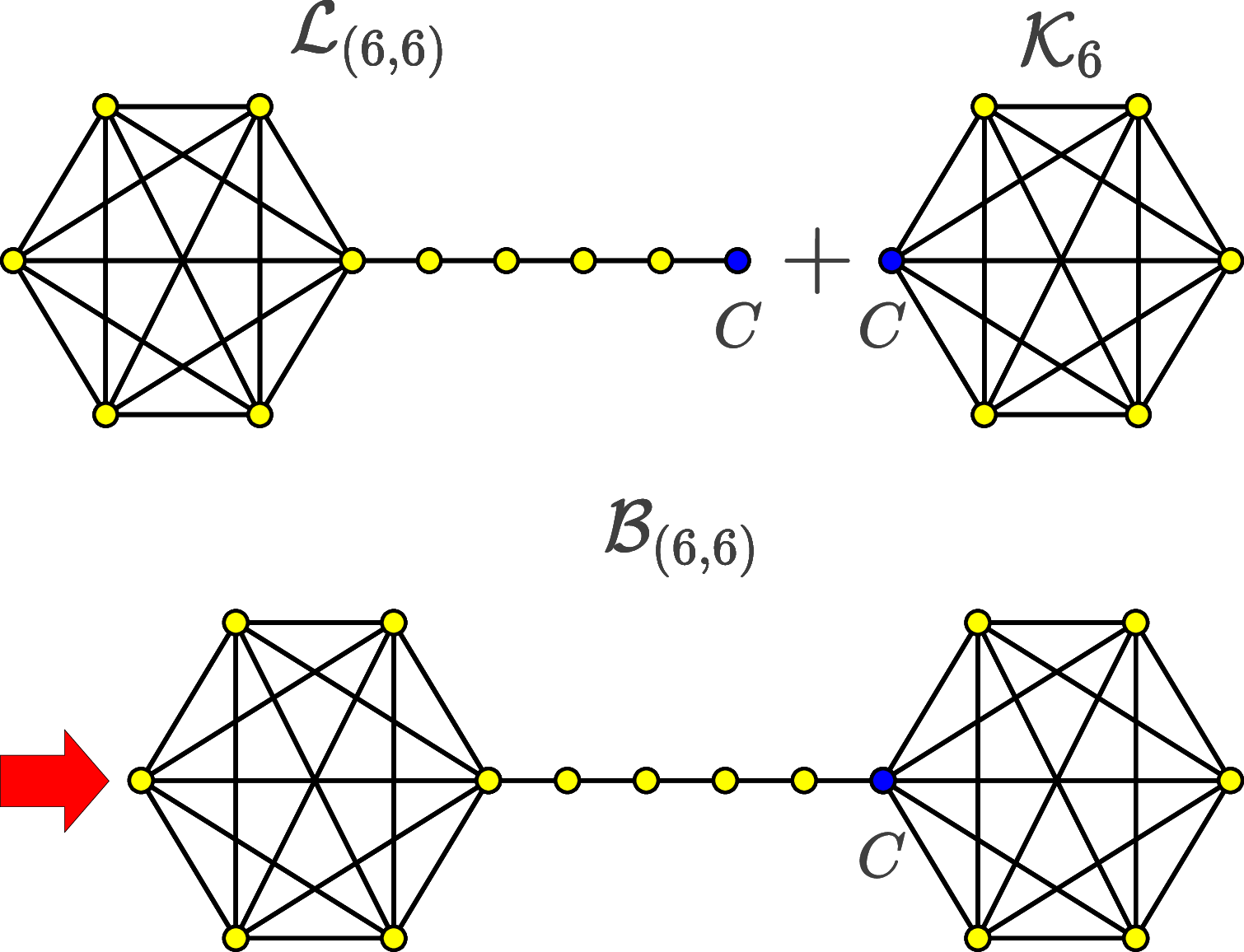}
\caption{A schematic illustration of the construction for the barbell graph $\mathcal{B}_{(6,6)}$. The barbell graph $\mathcal{B}_{(6,6)}$, shown on the right-hand side of the arrow, is formed by connecting a complete graph $\mathcal{K}_{6}$ to the endpoint $C$ of the lollipop graph $\mathcal{L}_{(6, 6)}$.
}
\label{fig:4}
\end{center}
\end{figure}

Furthermore, we use a barbell graph as an example to illustrate how the choice of the node $C$ can affect the transport efficiency of the graph $\mathcal{S}_2$. A barbell graph $\mathcal{B}_{(n_1,n_2)}$ is formed by connecting a complete graph $\mathcal{K}_{n_2}$ to the endpoint $C$ of a lollipop graph $\mathcal{L}_{(n_1,n_1)}$~\cite{aldous-fill-2014-book}. Fig.~\ref{fig:4} depicts the construction of a barbell graph $\mathcal{B}_{(6,6)}$. One can see that the total number of nodes in the barbell graph is $N^{\mathcal{B}_{(n_1,n_2)}}=2n_1+n_2-2$, while the total number of edges in the subgraphs $\mathcal{L}_{(n_1, n_1)}$ and $\mathcal{K}_{n_2}$ are $M_1 \sim (n_1)^2$ and $M_2 \sim (n_2)^2$, respectively.

Using Eqs.~(\ref{NK21}) and (\ref{NK31}), the MTT of endpoint $C$ in the lollipop graph $\mathcal{L}_{(n_1,n_1)}$ is easily computed to be $$\left\langle T^{\mathcal{L}_{(n_1,n_1)}}_C \right\rangle \sim (n_1)^3,$$ and the GFPT is $$\left\langle T^{\mathcal{L}_{(n_1,n_1)}} \right\rangle \sim (n_1)^2.$$ Similarly, let $n_1 = (n_2)^\lambda$, where $\lambda \leq 1$, then we have $$N^{\mathcal{B}_{(n_1,n_2)}} \sim n_2.$$ Using Eq.~(\ref{NK26}), we obtain the scaling of the GFPT in the barbell graph as
\begin{alignat}{1}\label{NK32}
\left\langle T^{\mathcal{B}_{(n_1,n_2)}} \right\rangle &\sim  (n_2)^{2\lambda}+n_2+\frac{(n_2)^2}{(n_2)^{2\lambda}+(n_2)^2}\times (n_2)^{3\lambda}
\nonumber\\ & \sim N^{\mathcal{B}_{(n_1,n_2)}}+(N^{\mathcal{B}_{(n_1,n_2)}})^{3\lambda}
\nonumber\\ & \sim (N^{\mathcal{B}_{(n_1,n_2)}})^{\alpha}\,,
\end{alignat}
where $\alpha \in [1, 3]$. In particular, when $\lambda=1$, we find that $\left\langle T^{\mathcal{B}_{(n_1,n_2)}} \right\rangle \sim (N^{\mathcal{B}_{(n_1,n_2)}})^{3}$, indicating that the transport efficiency is slower than that of both subgraphs, primarily due to the lower transport efficiency of the chosen node $C$.

%%%%%%%%%%%%%%%%%%%%%%%%%%%%%%%%%%%%%%%%%%%%%%%%%%%%%%%%%
% Figure  01
%%%%%%%%%%%%%%%%%%%%%%%%%%%%%%%%%%%%%%%%%%%%%%%%%%%%%%%%%%

Note that for a general network with $\text{GFPT} \sim N^\alpha$, the GFPT scaling exponent $\alpha$ is bounded by $1 \leq \alpha \leq 3$~\cite{ShengZhang-2019-IEEE-TIT, Tejedor-Benichou-Voituriez2009-PRE}. Eq.~(\ref{NK32}) shows that we can create a network with an arbitrary predefined GFPT scaling exponent by employing a barbell graph and choosing an appropriate value for $\lambda$.

Overall, this section presents a general approach for controlling the transport efficiency of networks, illustrated through examples of lollipop and barbell graphs. The findings discussed here are valuable for chemists and materials scientists in designing polymers and noncrystalline solids with specified anomalous diffusion and transport properties.

\section{Conclusion} \label{sec:Conclusion}
In this work, we aim to construct networks with different transport efficiencies by examining unbiased random walks on networks obtained by vertex merging operation. We analyze several key indicators: the MFPT between any two nodes, the MTT of individual nodes, and the GFPT of the network. Our findings indicate that these quantities are intrinsically linked to the first-passage properties of the corresponding subgraphs. Exact formulas describe the relations between these quantities in this class of networks and the related quantities in the subgraphs. GFPT is an important measure of information transport efficiency and stochastic search efficiency in networks. We propose a general method for controlling the transport (search) efficiency of the network. This involves selecting a suitable node and adjusting the growth of the number of nodes in the subgraph, enabling the construction of networks with varying transport efficiencies. As a practical example, we accurately compute the GFPT of the lollipop graph, thereby verifying our conclusions. Moreover, we demonstrate that networks with any GFPT scaling exponent $\alpha \in [1, 3]$ can be achieved within barbell graphs. Therefore, by selecting an appropriate node, we can design networks that fulfill specific practical requirements for transport (search) efficiency. The conclusions obtained here not only contribute to the design of efficient wireless networks and radar antennas but also offer valuable guidance for chemists and material scientists in creating polymers and noncrystalline solids with predetermined diffusion properties and transport efficiencies based on existing structures.

%The conclusions obtained here offer valuable guidance for chemists and material scientists in designing polymers and noncrystalline solids with predetermined diffusion properties and transport efficiencies based on existing structures.

\begin{IEEEbiography}[{\includegraphics[width=1in,height=1.25in,clip,keepaspectratio]{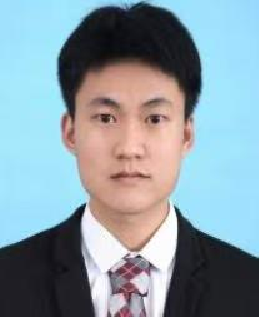}}]{Zhenhua Yuan}
is a Ph.D. candidate in the School of Mathematics and Information Science at Guangzhou University, Guangzhou, China. He received a M.S. degree in Mathematics and Information Science from the same university in 2023. His research interests lie in complex networks, random walks, and graph data mining.
\end{IEEEbiography}

\begin{IEEEbiography}[{\includegraphics[width=1in,height=1.25in,clip,keepaspectratio]{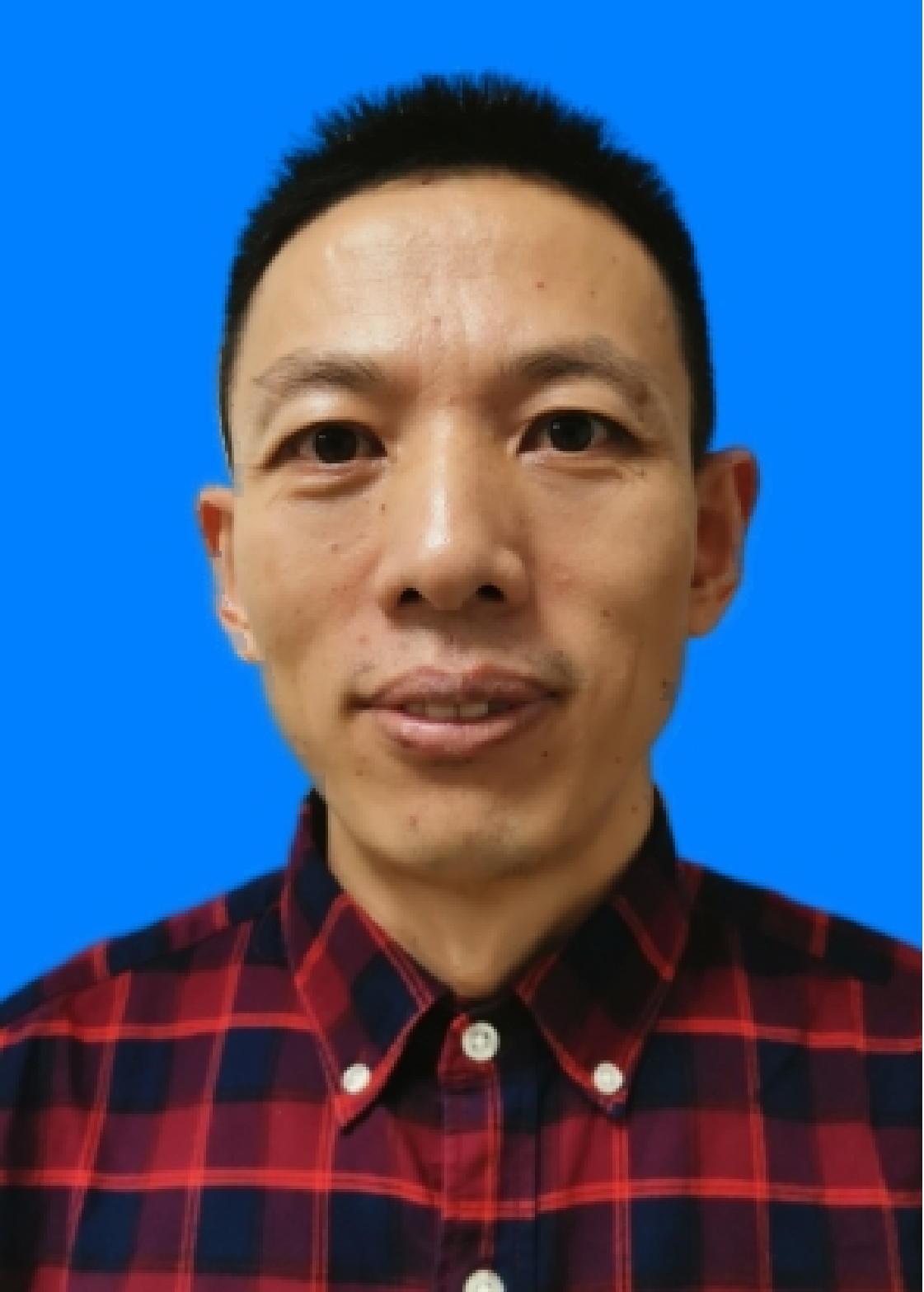}}]{Junhao Peng}
received the Ph.D. degree from the Beijing University of Posts and Telecommunications, China, in 2008. Currently he is a professor in School of Mathematics and Information Science, Guangzhou University, China. His research interests include network security and structural and dynamical properties of complex networks.
\end{IEEEbiography}

\begin{IEEEbiography}[{\includegraphics[width=1in,height=1.25in,clip,keepaspectratio]{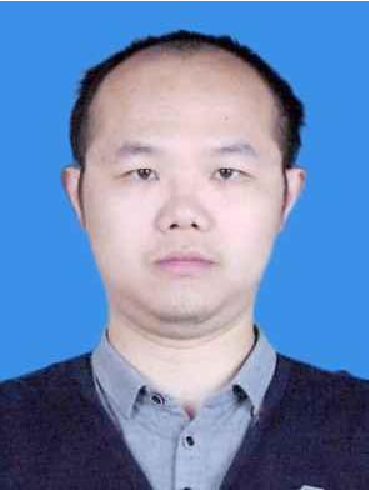}}]{Long Gao}
is currently conducting postdoctoral research at the School of Mathematics and Information Science, Guangzhou University, Guangzhou, China. He has authored or co-authored more than 10 peer-reviewed papers in venues including PRE, CHAOS, JSTAT, and FRACTS. His research interests include hyper-graph, random walks, complex networks, and random search.
\end{IEEEbiography}

\vfill

\end{document}